\documentclass[10pt,aps,prd,twocolumn,superscriptaddress,floatfix,amsmath,amssymb,amsfonts,longbibliography,nofootinbib]{revtex4-2}

\usepackage[utf8]{inputenc}   
\usepackage[T1]{fontenc}      
\usepackage{hyperref}         

\usepackage{graphicx}
\usepackage{amsmath, amssymb, amsfonts}
\usepackage{bm}
\usepackage{color}

\hypersetup{
    colorlinks = true,
    linkcolor = blue,
    citecolor = blue,
    urlcolor  = magenta
}

\begin{document}

\title{Spinning into Quantum Geometry: Dirac and Wheeler--DeWitt Dynamics from Stochastic Helicity}

\author{Partha Nandi}
\email{partha.nandi@nithecs.ac.za}
\affiliation{Department of Physics, Stellenbosch University, Stellenbosch 7600, South Africa}
\affiliation{National Institute for Theoretical and Computational Sciences (NITheCS), Stellenbosch 7604, South Africa}

\author{Partha Ghose}
\email{partha.ghose@gmail.com}
\affiliation{Tagore Centre for Natural Sciences and Philosophy, Rabindra Tirtha, New Town, Kolkata 700156, India}

\author{Francesco Petruccione}
\email{petruccione@sun.ac.za}
\affiliation{National Institute for Theoretical and Computational Sciences (NITheCS), Stellenbosch 7604, South Africa}
\affiliation{School of Data Science and Computational Thinking, Stellenbosch University, Stellenbosch 7600, South Africa}

\date{August 2025}

\begin{abstract}
Spin networks in loop quantum gravity provide a kinematical picture of quantum geometry but lack a natural mechanism for dynamical Dirac-type evolution, while the Wheeler–DeWitt equation typically enters only as an imposed constraint. We propose a stochastic framework in which each spin-network edge carries helicity-resolved amplitudes—two-state internal labels that undergo Poisson-driven flips. The resulting coupled master equations, after analytic continuation and the introduction of a fundamental length scale, generate Dirac-type dynamics on discrete geometry. At long times, the same process relaxes to helicity-symmetric equilibrium states, which are shown to satisfy a Wheeler–DeWitt-type condition. In this way, both quantum evolution and the gravitational constraint emerge within a single probabilistic framework. Our approach thus provides a background-independent and stochastic route to quantum geometry, offering an alternative to canonical quantization and a fresh perspective on the problem of time.
\end{abstract}

\maketitle

\section{Introduction}

The search for a consistent theory of quantum gravity confronts a deep conceptual divide.
Quantum mechanics describes matter with unrivaled precision but presupposes a fixed spacetime background, whereas general relativity treats spacetime itself as a dynamical entity shaped by energy and momentum.
Reconciling these views requires a reformulation of both geometry and dynamics at the most fundamental level.

A promising perspective is that quantum behavior itself may be emergent—the manifestation of a deeper stochastic substratum underlying spacetime and matter.
Stochastic quantization \cite{Lindgren:2019tdd}, introduced by F\'enyes~\cite{fenyes1966}, Kershaw~\cite{kershaw1966}, and Nelson~\cite{nelson1966}, provides a natural framework in which quantum amplitudes arise from probabilistic dynamics governed by diffusion-like processes.
In such a setting, both spacetime and matter fields can be viewed as collective excitations of an underlying stochastic substrate rather than as fundamental entities.
Within this picture, a stochastic formulation offers a natural way to reintroduce dynamical evolution without presupposing an external classical time parameter~\cite{ambjorn2009}, thereby addressing the ``problem of time'' that plagues canonical quantum gravity \cite{Unruh:1989db}.

Our central aim in this paper is to develop a \emph{stochastic helicity framework for spin networks} in which both Dirac-type evolution and Wheeler--DeWitt-type constraints~\cite{arnowitt1962} emerge dynamically from a single probabilistic process.
A distinctive feature of this approach is that the notion of \emph{time} itself arises as an intrinsic ordering parameter governing stochastic helicity transitions.
Thus, relativistic propagation, gravitational timelessness, and temporal emergence are shown to be complementary aspects of one underlying stochastic dynamics~\cite{Nandi:2025qrk}.
The main focus of this work is therefore the construction and analysis of this stochastic mechanism and its interpretation as a possible microscopic origin of geometric and dynamical structures familiar from quantum gravity.

Our stochastic helicity framework does not rely on breaking spacetime covariance in the conventional sense, because the evolution parameter $\tau$ serves as an intrinsic ordering parameter for the stochastic dynamics, providing a sequence of probabilistic transitions along the spin network.  
In this sense, our approach is complementary to recent developments in Lorentz-covariant quantum dynamics~\cite{Nandi:2023tfq}, where one of us formulated a framework that preserves covariance by modifying the standard Heisenberg algebra, without requiring time and space to be treated on equal footing.  
Moreover, recent studies suggest that stochasticity itself can induce noncommutative geometric features of spacetime~\cite{Arzano:2025, Nandi:2025qyj}, pointing to a deeper connection between randomness, covariance, and the emergent structure of quantum geometry.

Among the leading nonperturbative approaches to quantum gravity is loop quantum gravity (LQG), in which states of geometry are described by spin networks—discrete graphs labeled by group representations encoding areas and volumes~\cite{ashtekar1986}.
By combining stochastic quantization with spin-network states, we explore how a probabilistic substratum can generate a consistent quantum geometry and provide new insights into the emergence of spacetime and time itself.
Although the present work does not attempt a full quantization of gravity, it identifies a concrete stochastic mechanism through which key structures of canonical quantum gravity—Dirac dynamics, Hamiltonian constraints, and temporal ordering—arise naturally from a unified probabilistic origin.

In stochastic formulations of relativistic quantum theory~\cite{Breuer:1991ve,PhysRevA.56.2334,Breuer:2007juk}, Poisson-driven processes with internal state reversals are known to yield relativistic wave equations.
A classic example is the work of Gaveau, Jacobson, Kac, and Schulman~\cite{gaveau1984}, who showed that two coupled master equations for forward and backward movers map to the Dirac equation in the Weyl representation through analytic continuation.
We extend this insight to spin-network states in LQG by constructing a discrete stochastic model in which each edge carries helicity-resolved amplitudes evolving through coupled first-order equations.
This dynamics includes random helicity flips and directional propagation; after analytic continuation it gives rise to a Dirac-type evolution, while at long times it relaxes to helicity-symmetric equilibrium states satisfying Wheeler--DeWitt-type conditions.
In this picture, the stochastic parameter acts as an intrinsic ``clock'' ordering the transitions between helicity states—time arises dynamically from within the process itself.

In conventional field theory, helicity refers to the projection of spin onto momentum, well-defined in continuous spacetime.
In the background-independent setting of spin networks, no such notion of momentum exists.
Here ``helicity'' is employed in a generalized, operational sense: the amplitudes $\Psi^\pm_e$ represent internal two-state degrees of freedom associated with oriented edges, analogous to right- and left-moving components in the Weyl representation of the Dirac equation~\cite{Yordanov:2024}.
These states undergo stochastic transitions governed by a Poisson process, leading to parity-sensitive dynamics.
Upon analytic continuation, the components combine into a Dirac spinor, and their interchange under edge reversal mimics helicity flips in the relativistic setting.
In this sense, helicity transitions act as discrete geometric deformations of the spin network, encoding how local orientation, connectivity, and temporal ordering evolve stochastically.

Despite the success of canonical approaches such as the ADM formalism~\cite{arnowitt1962} and the Ashtekar variables~\cite{ashtekar1986}, major challenges persist, including the implementation of the Hamiltonian constraint, the recovery of semiclassical spacetime, and the coupling to matter fields.
The stochastic helicity framework proposed here offers an alternative, background-independent route in which geometric, dynamical, and temporal aspects emerge from a unified probabilistic process.
By emphasizing the stochastic origin of Dirac evolution and Wheeler--DeWitt-type equilibrium, the present work aims to clarify how the structural ingredients of canonical quantum gravity could arise from non-deterministic microscopic dynamics.

The structure of the paper is as follows.
Section~2 develops the stochastic helicity framework on spin networks.
Section~3 shows how Dirac-type evolution emerges.
Section~4 demonstrates that Wheeler--DeWitt-type conditions arise as equilibrium states.
Section~5 discusses couplings to matter and extensions to spin foams.
Section~6 addresses the emergence and dissolution of time.
Section~7 concludes.

\section{Stochastic Helicity Dynamics on Spin Networks}\label{sec:stochastic_dynamics}

Let $\Gamma$ be a graph embedded in a 3-manifold. Each directed edge $e$
carries two complex amplitudes, $\Psi^+_e(\tau)$ and $\Psi^-_e(\tau)$,
corresponding to positive and negative helicity states. Their dynamics is
modeled as a continuous-time Markov process \cite{Breuer:2007juk} with two contributions:

\begin{enumerate}
\item \emph{Helicity flips.} With Poisson rate $\lambda$, the helicity label
$+\leftrightarrow -$ flips. Over a short interval $\Delta\tau$, the amplitudes update as
\begin{eqnarray}
\Psi^+_e(\tau+\Delta \tau) &=& (1-\lambda \Delta \tau)\,\Psi^+_e(\tau) + \lambda \Delta \tau\,\Psi^-_e(\tau), \nonumber \\
\Psi^-_e(\tau+\Delta \tau) &=& (1-\lambda \Delta \tau)\,\Psi^-_e(\tau) + \lambda \Delta \tau\,\Psi^+_e(\tau). 
\end{eqnarray}
This linear mixing mirrors the standard master-equation update for a Poisson
process, but is applied directly to probability amplitudes rather than
classical probabilities. In the continuum limit $\Delta \tau \to 0$, one obtains
\begin{eqnarray}
\frac{d}{d\tau}\Psi^+_e &=& -\lambda\bigl(\Psi^+_e - \Psi^-_e\bigr), \nonumber \\
\frac{d}{d\tau}\Psi^-_e &=& -\lambda\bigl(\Psi^-_e - \Psi^+_e\bigr).
\end{eqnarray}

This term drives the two helicities toward equilibrium $\Psi^+_e=\Psi^-_e$, 
reflecting detailed balance.

\item \emph{Transport.} Amplitudes also propagate across adjacent nodes, 
encoded by terms $T_\pm(e)$ built from adjacency relations or intertwiners.
\end{enumerate}

Combining these effects, the stochastic evolution is

\begin{eqnarray}
\frac{d}{d\tau}\Psi^+_e &=& -\lambda \big(\Psi^+_e - \Psi^-_e\big) +T_+(e), \nonumber \\
\frac{d}{d\tau}\Psi^-_e &=& -\lambda \big(\Psi^-_e - \Psi^+_e\big) + T_-(e).
\label{eq:helicity-eqs}
\end{eqnarray}

This formulation shows that helicity is not microscopically conserved, 
but the dynamics drives the system toward helicity-symmetric equilibrium.

\subsection*{On Time Ordering in Spin Network Dynamics}

In background-independent quantum gravity, there is no external global time parameter. 
Spin networks describe ``instants'' of quantum geometry—solutions to Gauss and 
diffeomorphism constraints—but not dynamics. Spin foams provide histories of spin networks, 
where causal order is encoded combinatorially in the 2-complex structure.

In our stochastic framework, the parameter $\tau$ in eqn.~(\ref{eq:helicity-eqs}) is not 
geometric but a \emph{stochastic time}, recording the succession of probabilistic transitions. 
It provides an intrinsic ordering for the dynamics without introducing background time.

\section{From Stochastic Helicity Dynamics to Dirac-Type Evolution}
\label{sec:DiracDynamics}

To recover a quantum-like dynamics, we follow the analytic-continuation 
procedure introduced by Gaveau, Jacobson, Kac, and Schulman~\cite{gaveau1984}, 
who showed that coupled stochastic transport equations can be transformed 
into a Dirac equation through a continuation to imaginary time and a phase rotation. 
In our context, the same mechanism appears naturally in the evolution of 
helicity-resolved amplitudes on a spin network.

\medskip

We introduce a fundamental length scale $\ell$, representing the discrete spacing 
between adjacent vertices along a spin-network edge. Physically, $\ell$ may be 
identified with the Planck length—the minimal spacing in loop-quantum-gravity geometry \cite{rovelli2004}. 
Define the spinor-like object
\begin{equation}
\Psi_e =
\left(
\begin{array}{c}
\Psi^+_e \\
\Psi^-_e
\end{array}
\right).
\end{equation}
whose two components correspond to the helicity-resolved amplitudes propagating 
along the oriented edge~$e$.

The \emph{helicity-flip operator} $F_e$ acts by interchanging the two components,
\begin{equation}
F_e \Psi_e =
\left(
\begin{array}{c}
\Psi^-_e \\
\Psi^+_e
\end{array}
\right),
\end{equation}

\noindent
The coupled stochastic equations (\ref{eq:helicity-eqs}) can therefore 
be written compactly as
\begin{equation}
\frac{d}{d\tau} \Psi_e = -\lambda \left(\mathbb{I} - F_e\right)\Psi_e + \mathcal{T}_e,
\label{v}
\end{equation}
where $\mathbb{I}$ is the $2\times2$ identity matrix and
\begin{equation}
\mathcal{T}_e=\left(
\begin{array}{c}
T_{+}(e) \\[2mm]
T_{-}(e)
\end{array}
\right) =\frac{v}{\ell}
\left(
\begin{array}{c}
\Psi^+_{e}(x-\Delta x) - \Psi^+_{e}(x) \\[2mm]
\Psi^-_{e}(x+\Delta x) - \Psi^-_{e}(x)
\end{array}
\right).
\end{equation}
which is a finite-difference operator acting on the amplitudes with $\Delta x \sim \ell$. Here, $x$ is \emph{not} a spatial coordinate in spacetime; it is a discrete label for vertex position along an edge. The difference $x+\ell$ means “the next vertex along the oriented edge,” and $x-\ell$ means “the previous vertex.” 
$\ell$ is therefore the discrete spacing between adjacent vertices along the edge, while $v$ is the propagation speed or transport coefficient along the edge $e$.


\begin{figure}[t]
    \centering
    \includegraphics[width=0.45\textwidth]{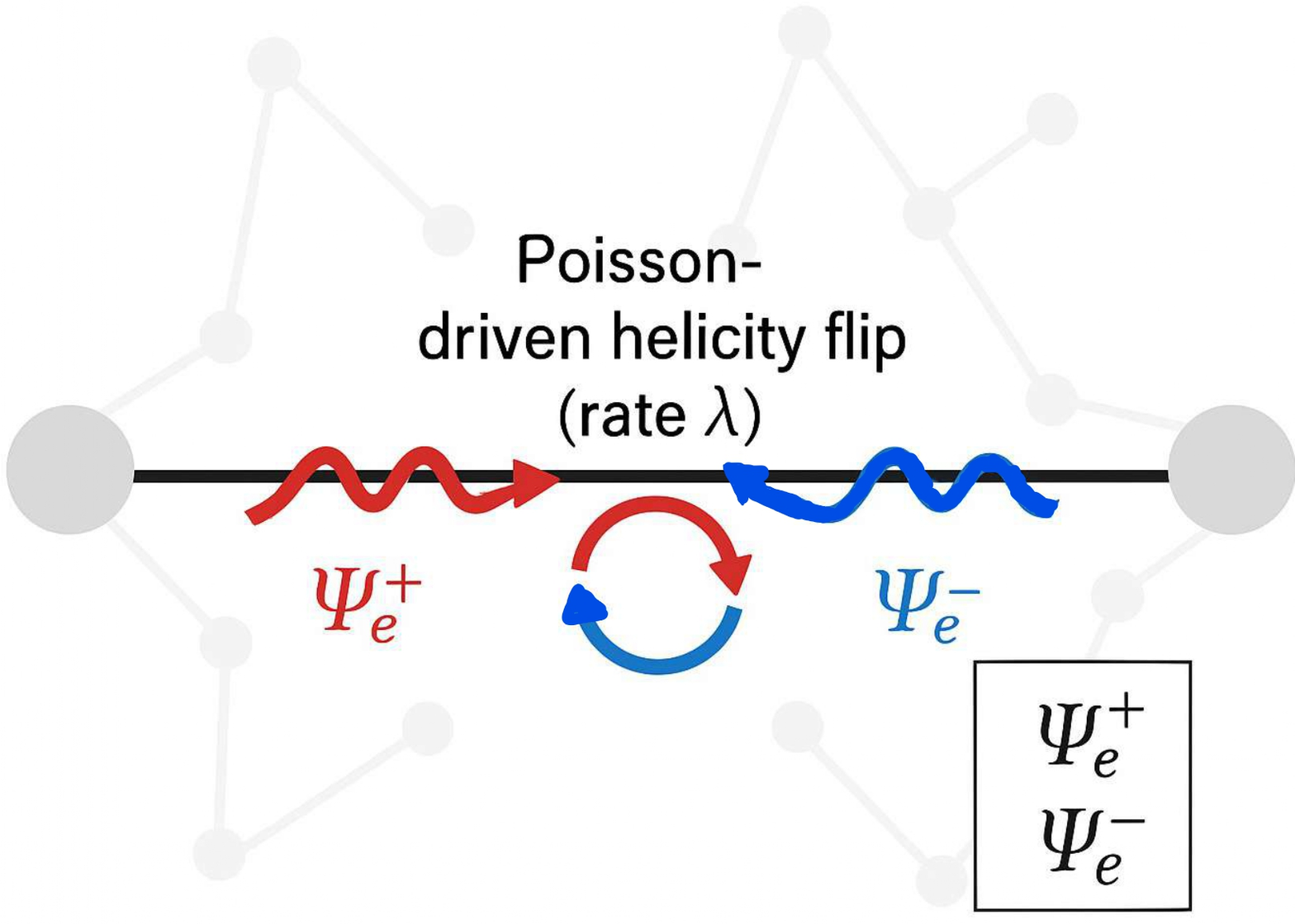}
    \caption{Schematic diagram showing helicity-resolved amplitudes $\Psi^+_e$ 
    and $\Psi^-_e$ propagating along a spin-network edge~$e$, with Poisson-driven 
    helicity flip transitions.}
    \label{fig:helicity_flip}
\end{figure}

\medskip

In fact, expanding $\Psi^{\pm}_{e}(x \mp \ell)$ in powers of $\ell$ gives
\begin{equation}
\Psi^{\pm}_{e}(x \mp \ell,\tau) = \Psi^{\pm}_{e}(x,\tau) \mp \ell \, \partial_x \Psi^{\pm}_{e}(x,\tau) + \frac{\ell^2}{2} \, \partial_x^2 \Psi^{\pm}_{e}(x,\tau) + \cdots.
\end{equation}
From this expansion, it is straightforward to see that, to leading order in $\ell$, the transport operator reduces to

\begin{equation} \mathcal{T}_e \;\sim\;v \left( \begin{array}{c} -\partial_x \Psi^+(x,\tau) \\[2mm]  \partial_x \Psi^-(x,\tau) \end{array} \right). \end{equation}

This can equivalently be written in matrix form as
\begin{equation}
\mathcal{T} \longrightarrow -\sigma^3 \, v\partial_x \psi(x,\tau), 
\qquad \sigma^3 = \left(
\begin{array}{cc}
1 & 0 \\
0 & -1
\end{array}
\right).
\end{equation}

Inserting the continuum-expanded result into the evolution equation~(\ref{v}), 
the amplitudes are now functions $\Psi_{e}(x,\tau)$ of both space and time. 
Accordingly, the total time derivative is replaced by a partial derivative, 
and we obtain
\begin{equation}\label{eq:discrete_continuum}
 \partial_\tau \Psi_e(x,\tau)  
= - \lambda \,(\mathbb{I}- \sigma^1) \Psi_e(x,\tau) 
-  v \, \sigma^3 \, \partial_x \Psi_e(x,\tau).
\end{equation}

The first term describes stochastic helicity flips with rate~$\lambda$, while 
the second encodes propagation along the edge.  
Equation~(\ref{eq:discrete_continuum}) is structurally identical to the 
classical telegrapher’s equation analyzed by Gaveau et al.~\cite{gaveau1984}.  
To connect with their construction, we now make the analytic continuation 
to imaginary time, through the anti Wick rotation $\tau\mapsto -i \tau$ , which converts the real, dissipative 
generator into an Hermitian one, producing oscillatory evolution.

Under this continuation (and recalling that $v=Lt_{\Delta \tau \rightarrow 0}\Delta x/\Delta \tau$ transforms as a velocity), 
we replace $\partial_\tau\!\to\! i\partial_\tau$ and $v\!\to\! i v$, obtaining
\begin{equation}
i\,\partial_\tau\Psi_e(x,\tau)
= -\lambda(\mathbb{I}-\sigma^1)\Psi_e(x,\tau)
- i v\,\sigma^3\,\partial_x\Psi_e(x,\tau).
\label{eq:quantum-like}
\end{equation}

\noindent
The term proportional to the identity, $-\lambda\mathbb{I}$, contributes only 
a global phase and may be removed by a transformation to a rotating frame 
$\Psi_e\mapsto \Psi^{'}_e= e^{i\lambda t}\Psi_e$.  
The physically relevant evolution equation then reduces to
\begin{equation}
i\hbar\,\partial_\tau\Psi^{'}_e(x,\tau)
= \lambda \hbar\,\sigma^1\Psi^{'}_e(x,\tau)
- i \hbar v\,\sigma^3\,\partial_x\Psi^{'}_e(x,\tau),
\label{eq:dirac-pre}
\end{equation}
where we have multiplied both sides by $\hbar$. This is precisely the Dirac-type structure obtained by Gaveau \textit{et al.} after analytic continuation. It may be noted that the above form is nothing but the one-dimensional Dirac equation in the Weyl representation, as also discussed in their work. With the identifications  
\[
c \leftrightarrow v,\qquad 
\alpha \leftrightarrow \sigma^3,\qquad 
\beta \leftrightarrow \sigma^1,\qquad 
mc^2 \leftrightarrow \lambda\hbar,
\]
the equation can be rewritten in a manifestly covariant form as

\begin{equation}
\big(i\hbar\gamma^\mu\partial_\mu - mc\big)\Psi'_e (x,\tau) = 0.
\label{hb}
\end{equation}
with $\gamma^0=\sigma_1$ and $\gamma^1=i\sigma_3$. The structure thus clearly exhibits the Lorentz-covariant Dirac like dynamics in $(1+1)$ dimensions, with $\Psi^{'}_e$ representing a two-component spinor field evolving under an effective mass generated by the coupling $\lambda$.
The stochastic flip rate $\lambda$ thus plays the role of an effective mass,
while the transport coefficient $v$ sets the propagation velocity.


The analytic continuation has therefore converted the dissipative stochastic dynamics into
an oscillatory, unitary one, generating a Dirac-type evolution for the helicity-resolved
amplitudes on each spin-network edge. The parameters $\lambda$ and $v$ determine the effective
mass and propagation velocity of this emergent relativistic field.

However, this Dirac-like dynamics describes only the \emph{non-equilibrium regime}, in which
helicity populations continue to fluctuate under stochastic flips.
Over long stochastic times, these flips drive the system toward \emph{helicity symmetry},
$\Psi_e^{+}\!\approx\!\Psi_e^{-}$, suppressing net helicity flow.
This relaxation represents the approach to equilibrium of the stochastic process.
As shown next, the stationary (equilibrium) sector of the same dynamics satisfies a
Wheeler--DeWitt--type constraint, providing the timeless limit of the theory.

\section{Wheeler--DeWitt-Type Condition from Stochastic Equilibrium}

The analytic continuation discussed in Sec.~3 revealed that the stochastic helicity process,
when viewed in the short-time regime, generates a coherent Dirac-type evolution.
However, the same underlying stochastic mechanism also admits a complementary
long-time behavior.  As the Poisson-driven helicity flips continue to act,
they gradually equalize the two helicity amplitudes, driving the system toward
helicity symmetry and statistical equilibrium.  In this limit, stochastic
transitions cease to produce net change, and the dynamics becomes stationary.

\medskip
Returning to the spinor form of the stochastic equation,
\begin{equation}
\frac{d}{d\tau}\Psi_e = -\lambda (\mathbb{I}-F_e)\Psi_e + T_e ,
\label{eq:stochastic-eq}
\end{equation}
the approach to equilibrium is characterized by $d\Psi_e/d\tau \!\to\! 0$ and
$T_e\!\to\!0$.  The remaining condition,
\begin{equation}
(\mathbb{I}-F_e)\Psi_e = 0 ,
\label{eq:helicity-symmetry}
\end{equation}
selects states that are invariant under helicity reversal, $\Psi_e^{+}=\Psi_e^{-}$.
Equation~\ref{eq:helicity-symmetry} thus defines the
\emph{helicity-symmetric subspace} of the stochastic dynamics.

Summing the local operators $(\mathbb{I}-F_e)$ over all edges gives a global constraint operator
\begin{equation}
\hat{\mathcal H} = \hbar\lambda \sum_{e}(\mathbb{I}-F_e),
\label{eq:H-operator}
\end{equation}
so that equilibrium states satisfy
\begin{equation}
\hat{\mathcal H}\Psi = 0 .
\label{eq:WDW-constraint}
\end{equation}
Here $\lambda$ is the stochastic flip rate and $\hbar\lambda$ sets the characteristic
energy scale of the constraint operator.  Equation~\ref{eq:WDW-constraint} therefore
plays the role of a Wheeler--DeWitt--type condition on the stochastic spin-network
state, expressing vanishing ``total energy'' in the equilibrium limit.

\medskip

The Dirac and Wheeler--DeWitt equations arise as two complementary limits of a single
stochastic helicity dynamics: the Dirac equation governs coherent, unitary
propagation away from equilibrium, while the Wheeler--DeWitt condition describes the
stationary, timeless configuration reached once helicity symmetry is restored.

Importantly, the Wheeler--DeWitt condition is not an external postulate but emerges as
the \emph{fixed point} of the stochastic evolution.  Helicity-symmetric equilibrium
states belong to the kernel of $\hat{\mathcal H}$, corresponding to vanishing net
helicity flow---directly analogous to the vanishing of the Hamiltonian in canonical
quantum gravity.  In this way, Einstein-type constraints appear as conditions of
stochastic equilibration, providing a unified probabilistic framework that connects
Dirac-like dynamical evolution with gravitational timelessness.




\section{Coupling to Matter Fields and Spin Foam Extensions}

A natural question is how matter degrees of freedom may couple to the
helicity-resolved spin network dynamics. In conventional lattice gauge
theory, scalar and fermionic fields reside on vertices, while link
variables encode parallel transport across edges. Interactions are
introduced by combining vertex fields with edge variables in a way
that respects locality and internal symmetries. The same guiding
principle can be applied here: in our stochastic helicity framework,
edge amplitudes $\Psi^\pm(e)$ play the role of link variables, while
matter fields $\phi(v), \psi(v)$ are associated with the vertices
$v \in \Gamma$.

\subsection*{Scalar couplings}
For scalar matter, a minimal vertex–edge coupling is obtained by
summing the scalar field over the endpoints of an edge and weighting
the corresponding helicity amplitude. This yields a local source term
of the form
\begin{equation}
J_\pm(e; \phi) = g \sum_{v \in \partial e} \phi(v) \, \Psi_\pm(e),
\end{equation}
where $g$ is a coupling constant and $\partial e$ denotes the vertices
incident on edge $e$. Such couplings bias the stochastic helicity
evolution in a manner analogous to scalar–link couplings in lattice
gauge theory, effectively renormalizing the helicity dynamics through
local scalar fields.

\subsection*{Spinor couplings}

Fermionic matter on the spin network must respect the two-state helicity structure intrinsic to each edge.  
In our framework, helicity does not refer to spin projection along momentum, since no notion of particle momentum exists.  
Instead, it is an internal orientation degree of freedom associated with the edge, corresponding to the two helicity amplitudes $\Psi_e^\pm$ introduced in Section~\ref{sec:stochastic_dynamics}.  
This motivates the introduction of helicity-sensitive projectors
\begin{equation}
P_\pm = \frac{1}{2} \left( \mathbb{I} \pm \hat{n}_e \cdot \vec{\Sigma} \right),
\end{equation}
where $\vec{\Sigma}$ are spin generators acting on the internal spinor space, and $\hat{n}_e$ is the unit vector along the oriented edge $e$.  
The projectors select the components aligned or anti-aligned with the edge orientation and provide a discrete analogue of helicity projection in the absence of momentum.  

A generic vertex--edge interaction can then be written as
\begin{equation}
J_\pm(e;\psi) = g' \sum_{v \in \partial e} \bar{\psi}(v) \, \gamma^\mu P_\pm \psi(v) \, A_\mu(e),
\end{equation}
where $g'$ is a coupling constant, $\psi(v)$ is a vertex spinor, and $A_\mu(e)$ is an effective connection associated with the edge.  

The effective connection $A_\mu(e)$ is not an external field \cite{Alvarez-Gaume:1983ihn}; it emerges from the stochastic helicity dynamics along the edge as described by the telegrapher equation~(\ref{eq:discrete_continuum}).  
To leading order in the continuum expansion of finite-difference transport, the relative change of the positive-helicity amplitude in the direction of the negative-helicity amplitude defines
\begin{equation}
A_\mu(e) \approx i\, \Psi_e^{-\dagger} \nabla_\mu \Psi_e^+ + h.c.,
\end{equation}
which is Hermitian and generates unitary parallel transport in helicity space.  
Here, $\nabla_\mu \Psi^+$ represents the derivative along the edge, obtained either from the finite-difference transport operator $\mathcal{T}_e$ or its continuum limit.  

This construction mirrors lattice gauge-theory couplings, where spinor bilinears are transported across links.  
In our framework, the transport is intrinsically determined by the edge orientation $\hat{n}_e$ and the stochastic helicity dynamics.  
Fermionic degrees of freedom therefore propagate across the spin network in a helicity-dependent, orientation-sensitive manner, while preserving background independence and consistency with the telegrapher-based stochastic evolution introduced earlier.

\subsection*{Extension to Spin Foams}
The stochastic spin network admits a natural extension to spin foams,
viewed as histories of spin network transitions. In this setting,
helicity degrees of freedom $h_f = \pm 1$ can be assigned to faces of
the 2-complex. A transition amplitude between initial and final
spin networks can then be written as a helicity-resolved statistical
sum,
\begin{equation}
\mathcal{A}(s_i \rightarrow s_f) = 
\sum_{\{h_f\}} \prod_{f} w_f(h_f) \prod_{v} \mathcal{A}_v(h_{f \supset v}),
\end{equation}
where $w_f(h_f)$ are helicity-dependent face weights and
$\mathcal{A}_v$ are vertex amplitudes encoding helicity transition
rules. This parallels the continuum construction of
Gaveau, Jacobson, Kac, and Schulman~\cite{gaveau1984}, where coupled
master equations give rise to the Dirac equation via analytic
continuation. In the present context, helicity-resolved spin foam
amplitudes emerge as sums over discrete histories weighted by
stochastic transition rules.

\medskip

In this way, the introduction of matter couplings is not ad hoc but
guided by the same principles that underlie lattice gauge theory and
loop quantum gravity: vertex fields couple locally to edge variables,
and helicity projectors ensure consistency with the internal two-state
structure. This opens the possibility of describing both geometry and
matter within the same stochastic helicity framework.

\section{The Emergence and Dissolution of Time}
\label{sec:Time}

One of the central conceptual challenges in quantum gravity is the nature and role of time.
In canonical approaches such as the ADM formalism~\cite{ADM1962} and the Wheeler--DeWitt framework, time does not appear as an explicit variable.
Quantum dynamics is instead governed by the Hamiltonian constraint,
\begin{equation}
\hat{H}\Psi[\mathrm{geometry}] = 0,
\end{equation}
which implies that physical states of the universe are ``frozen.''
This leads to the well-known \emph{problem of time}~\cite{Kuchar1992, Isham1992}, where the apparent timelessness of the fundamental description must be reconciled with the experience of dynamical evolution.

In the stochastic helicity framework developed here, the notion of geometry is represented not by a spatial metric but by the relational data carried by the helicity-resolved spin-network amplitudes $\Psi_e^{\pm}(t)$ introduced in Sec.~\ref{sec:stochastic_dynamics}.
These amplitudes, associated with edges $e$ of the network, encode adjacency and local orientation, while the helicity-flip and transport operators $(\mathcal{F}_e,\mathcal{T}_e)$ govern transitions that generate effective curvature and connectivity.
The collective configuration $\{\Psi_e^{\pm}\}$ thus constitutes the stochastic analogue of spatial geometry, and its stationary correlations define the emergent geometric structure.

As discussed in Sec.~\ref{sec:stochastic_dynamics}, the amplitudes obey the stochastic evolution equations~(\ref{eq:stochastic-eq}), where Poisson-driven helicity flips and transport events generate probabilistic transitions along each edge.
After analytic continuation, this same dynamics gives rise to the Dirac-type equation~(\ref{hb}) of Sec.~\ref{sec:DiracDynamics}.
The parameter $t$ in these equations plays a role analogous to the time variable in nonequilibrium statistical mechanics~\cite{gaveau1984}: it serves as an internal ordering parameter that labels the succession of stochastic transitions, but it has no geometric meaning and introduces no preferred foliation.

The continued action of stochastic helicity flips drives the system toward equilibrium.
In the long-time limit of Eq.~(\ref{eq:stochastic-eq}), one finds $d\Psi_e^{\pm}/d\tau \!\to\! 0$, and the network becomes helicity-symmetric, satisfying the Wheeler--DeWitt--type condition~(\ref{eq:WDW-constraint}).
At this point the dependence on the auxiliary parameter $\tau$ disappears, and reparametrization invariance is dynamically restored.
Timelessness therefore arises as the equilibrium limit of the stochastic process, not as an externally imposed postulate.

While $\tau$ governs microscopic stochastic evolution, it is not part of spacetime itself.
Spacetime geometry appears only once the network has equilibrated, when stationary correlations among the helicity amplitudes form stable relational patterns that can be represented by effective geometric quantities such as adjacency, curvature, or causal order.
The equilibrium configuration thus defines a reparametrization-invariant sector whose relational data encode the emergent geometry.

In this view, stochastic helicity dynamics provides a natural mechanism for the \emph{emergence} of time as a pre-geometric ordering parameter away from equilibrium and for its \emph{dissolution} once equilibrium is reached.
Spacetime and its geometric notion of time then appear as large-scale manifestations of this underlying stochastic order.
Hence, what appears as the ``problem of time'' in canonical quantum gravity is reinterpreted here as a dynamical transition from microscopic stochastic evolution to macroscopic geometric stationarity.

\begin{table*}[t]
\centering
\resizebox{\textwidth}{!}{
\begin{tabular}{|l|p{2cm}|p{3.2cm}|p{3.2cm}|}
\hline
\textbf{Framework} & \textbf{ADM Formalism} & \textbf{Ashtekar Variables / LQG} & \textbf{Stochastic Helicity Model (This Work)} \\
\hline
Quantization Method & Canonical (metric variables) & Canonical (connection variables) & Stochastic evolution with analytic continuation \\
\hline
Constraint Structure & Non-polynomial Hamiltonian constraint & Polynomial constraints in SU(2) variables & Helicity-symmetric condition as Wheeler--DeWitt-type constraint \\
\hline
Degrees of Freedom & Spatial metric and conjugate momentum & SU(2) connections and densitized triads & Helicity-resolved amplitudes on spin network edges \\
\hline
Treatment of Time & Problematic (frozen formalism) & Time evolution via constraints or spin foams & First-order time evolution with emergent spatial derivatives \\
\hline
Matter Coupling & Technically complex & Admissible but often indirect & Direct coupling to helicity-resolved states \\
\hline
Geometrical Interpretation & Geometro\-dynamics & Loop quantum geometry & Stochastic geometry with internal helicity structure \\
\hline
\end{tabular}
}
\caption{Comparison of canonical approaches to quantum gravity with the stochastic helicity model proposed in this paper.}
\label{tab:comparison}
\end{table*}

\section{Conclusion}

In this work, we developed a stochastic helicity framework for spin networks that provides a background-independent route toward quantum geometry. 
Each edge of the network carries helicity-resolved amplitudes $\Psi_e^\pm$, evolving through Poisson-driven flips and transport processes.
By analytically continuing this stochastic dynamics, we demonstrated that a Dirac-type evolution equation emerges naturally on discrete geometry, while the approach to stochastic equilibrium restores helicity symmetry and yields the Wheeler--DeWitt-type constraint~(\ref{eq:WDW-constraint}). 
Both relativistic propagation and gravitational constraints therefore arise dynamically within a single probabilistic setting, rather than being externally imposed as in canonical quantization.

The gravitational content of the theory resides in the helicity amplitudes $\Psi_e^\pm$, which act as microscopic geometric variables analogous to connection or curvature degrees of freedom, while the spin-network connectivity specifies the discrete kinematic geometry. 
Their stochastic evolution captures quantum-geometric fluctuations, and the relaxation toward equilibrium dynamically enforces a Hamiltonian-like constraint—realizing classical consistency without reference to a predefined spacetime metric.

A central conceptual result concerns the nature of time. 
Here, time emerges not as a background coordinate but as an intrinsic ordering parameter governing the stochastic transitions of helicity amplitudes. 
This picture parallels the Page--Wootters framework~\cite{Page:1983uc}, in which time arises from conditional correlations within a stationary global state.
In the non-equilibrium regime, this internal parameter drives genuine dynamical evolution; as equilibrium is reached, helicity symmetry freezes the dynamics and the Wheeler--DeWitt-type constraint is recovered. 
Time therefore both arises and dissolves dynamically within the same stochastic process, reconciling evolution and timelessness within a single framework.
Since the parameter $\tau$ is intrinsic rather than geometric, background independence and covariance are preserved throughout.

These results suggest that the core structures of quantum gravity—Dirac dynamics, Hamiltonian constraints, and the emergence of temporality—can be viewed as statistical consequences of an underlying stochastic substratum. 
Spacetime geometry and causal structure appear as large-scale correlations among microscopic helicity degrees of freedom, rather than as quantized fields on a preexisting manifold.
This reformulation provides a unifying perspective that connects dynamical and constraint-based formulations of quantum gravity within a single probabilistic framework.

Future extensions of this approach include: (i) constructing higher-dimensional generalizations and establishing the discrete-to-continuum correspondence linking stochastic correlations to effective spacetime geometry; (ii) introducing matter couplings and analyzing renormalization under coarse-graining; and (iii) formulating a stochastic spin-foam path integral that may bridge equilibrium dynamics with loop quantum cosmology. 
Such developments could clarify how isotropization, early-universe evolution, and a cosmological arrow of time emerge from stochastic equilibration.

In summary, stochastic helicity dynamics unifies Dirac-type evolution,
Wheeler--DeWitt-type constraints, and a dynamical treatment of time
within a single background-independent framework.
The helicity amplitudes and their stochastic correlations constitute
microscopic gravitational degrees of freedom whose equilibrium structure
encodes the emergence of classical spacetime geometry.
This provides a statistical origin for both spacetime and gravitational
constraints, linking quantum dynamics and timelessness within one
probabilistic formulation.

\section{Acknowledgments}
PN acknowledges support from the Rector’s Postdoctoral Fellowship Program (RPFP) at Stellenbosch University.  PG thanks the Director and staff of
IIT Mandi for their warm support and hospitality during
his visit, when this work was completed.




\begin{thebibliography}{23}%
\makeatletter
\providecommand \@ifxundefined [1]{%
 \@ifx{#1\undefined}
}%
\providecommand \@ifnum [1]{%
 \ifnum #1\expandafter \@firstoftwo
 \else \expandafter \@secondoftwo
 \fi
}%
\providecommand \@ifx [1]{%
 \ifx #1\expandafter \@firstoftwo
 \else \expandafter \@secondoftwo
 \fi
}%
\providecommand \natexlab [1]{#1}%
\providecommand \enquote  [1]{``#1''}%
\providecommand \bibnamefont  [1]{#1}%
\providecommand \bibfnamefont [1]{#1}%
\providecommand \citenamefont [1]{#1}%
\providecommand \href@noop [0]{\@secondoftwo}%
\providecommand \href [0]{\begingroup \@sanitize@url \@href}%
\providecommand \@href[1]{\@@startlink{#1}\@@href}%
\providecommand \@@href[1]{\endgroup#1\@@endlink}%
\providecommand \@sanitize@url [0]{\catcode `\\12\catcode `\$12\catcode `\&12\catcode `\#12\catcode `\^12\catcode `\_12\catcode `\%12\relax}%
\providecommand \@@startlink[1]{}%
\providecommand \@@endlink[0]{}%
\providecommand \url  [0]{\begingroup\@sanitize@url \@url }%
\providecommand \@url [1]{\endgroup\@href {#1}{\urlprefix }}%
\providecommand \urlprefix  [0]{URL }%
\providecommand \Eprint [0]{\href }%
\providecommand \doibase [0]{https://doi.org/}%
\providecommand \selectlanguage [0]{\@gobble}%
\providecommand \bibinfo  [0]{\@secondoftwo}%
\providecommand \bibfield  [0]{\@secondoftwo}%
\providecommand \translation [1]{[#1]}%
\providecommand \BibitemOpen [0]{}%
\providecommand \bibitemStop [0]{}%
\providecommand \bibitemNoStop [0]{.\EOS\space}%
\providecommand \EOS [0]{\spacefactor3000\relax}%
\providecommand \BibitemShut  [1]{\csname bibitem#1\endcsname}%
\let\auto@bib@innerbib\@empty
\bibitem [{\citenamefont {Lindgren}\ and\ \citenamefont {Liukkonen}(2019)}]{Lindgren:2019tdd}%
  \BibitemOpen
  \bibfield  {author} {\bibinfo {author} {\bibfnamefont {J.}~\bibnamefont {Lindgren}}\ and\ \bibinfo {author} {\bibfnamefont {J.}~\bibnamefont {Liukkonen}},\ }\href@noop {} {\bibfield  {journal} {\bibinfo  {journal} {Scientific Reports}\ }\textbf {\bibinfo {volume} {9}},\ \bibinfo {pages} {19984} (\bibinfo {year} {2019})}\BibitemShut {NoStop}%
\bibitem [{\citenamefont {Fényes}(1966)}]{fenyes1966}%
  \BibitemOpen
  \bibfield  {author} {\bibinfo {author} {\bibfnamefont {I.}~\bibnamefont {Fényes}},\ }\href@noop {} {\bibfield  {journal} {\bibinfo  {journal} {Zeitschrift für Physik}\ }\textbf {\bibinfo {volume} {132}},\ \bibinfo {pages} {81} (\bibinfo {year} {1966})}\BibitemShut {NoStop}%
\bibitem [{\citenamefont {Kershaw}(1966)}]{kershaw1966}%
  \BibitemOpen
  \bibfield  {author} {\bibinfo {author} {\bibfnamefont {D.}~\bibnamefont {Kershaw}},\ }\href@noop {} {\bibfield  {journal} {\bibinfo  {journal} {Journal of Physics A}\ }\textbf {\bibinfo {volume} {4}},\ \bibinfo {pages} {256} (\bibinfo {year} {1966})}\BibitemShut {NoStop}%
\bibitem [{\citenamefont {Nelson}(1966)}]{nelson1966}%
  \BibitemOpen
  \bibfield  {author} {\bibinfo {author} {\bibfnamefont {E.}~\bibnamefont {Nelson}},\ }\href@noop {} {\bibfield  {journal} {\bibinfo  {journal} {Physical Review}\ }\textbf {\bibinfo {volume} {150}},\ \bibinfo {pages} {1079} (\bibinfo {year} {1966})}\BibitemShut {NoStop}%
\bibitem [{\citenamefont {Ambj{\o}rn}\ \emph {et~al.}(2009)\citenamefont {Ambj{\o}rn}, \citenamefont {Loll}, \citenamefont {Westra},\ and\ \citenamefont {Zohren}}]{ambjorn2009}%
  \BibitemOpen
  \bibfield  {author} {\bibinfo {author} {\bibfnamefont {J.}~\bibnamefont {Ambj{\o}rn}}, \bibinfo {author} {\bibfnamefont {R.}~\bibnamefont {Loll}}, \bibinfo {author} {\bibfnamefont {W.}~\bibnamefont {Westra}},\ and\ \bibinfo {author} {\bibfnamefont {S.}~\bibnamefont {Zohren}},\ }\href@noop {} {\bibfield  {journal} {\bibinfo  {journal} {Physics Letters B}\ }\textbf {\bibinfo {volume} {678}},\ \bibinfo {pages} {1} (\bibinfo {year} {2009})}\BibitemShut {NoStop}%
\bibitem [{\citenamefont {Unruh}\ and\ \citenamefont {Wald}(1989)}]{Unruh:1989db}%
  \BibitemOpen
  \bibfield  {author} {\bibinfo {author} {\bibfnamefont {W.~G.}\ \bibnamefont {Unruh}}\ and\ \bibinfo {author} {\bibfnamefont {R.~M.}\ \bibnamefont {Wald}},\ }\href@noop {} {\bibfield  {journal} {\bibinfo  {journal} {Physical Review D}\ }\textbf {\bibinfo {volume} {40}},\ \bibinfo {pages} {2598} (\bibinfo {year} {1989})}\BibitemShut {NoStop}%
\bibitem [{\citenamefont {Arnowitt}\ \emph {et~al.}(1962{\natexlab{a}})\citenamefont {Arnowitt}, \citenamefont {Deser},\ and\ \citenamefont {Misner}}]{arnowitt1962}%
  \BibitemOpen
  \bibfield  {author} {\bibinfo {author} {\bibfnamefont {R.}~\bibnamefont {Arnowitt}}, \bibinfo {author} {\bibfnamefont {S.}~\bibnamefont {Deser}},\ and\ \bibinfo {author} {\bibfnamefont {C.~W.}\ \bibnamefont {Misner}},\ }in\ \href@noop {} {\emph {\bibinfo {booktitle} {Gravitation: An Introduction to Current Research}}},\ \bibinfo {editor} {edited by\ \bibinfo {editor} {\bibfnamefont {L.}~\bibnamefont {Witten}}}\ (\bibinfo  {publisher} {Wiley},\ \bibinfo {address} {New York},\ \bibinfo {year} {1962})\ pp.\ \bibinfo {pages} {227--265}\BibitemShut {NoStop}%
\bibitem [{\citenamefont {Nandi}\ and\ \citenamefont {Ghose}(2025)}]{Nandi:2025qrk}%
  \BibitemOpen
  \bibfield  {author} {\bibinfo {author} {\bibfnamefont {P.}~\bibnamefont {Nandi}}\ and\ \bibinfo {author} {\bibfnamefont {P.}~\bibnamefont {Ghose}},\ }\href@noop {} {\bibfield  {journal} {\bibinfo  {journal} {arXiv preprint}\ } (\bibinfo {year} {2025})},\ \Eprint {https://arxiv.org/abs/2508.10190} {arXiv:2508.10190 [gr-qc]} \BibitemShut {NoStop}%
\bibitem [{\citenamefont {Nandi}\ and\ \citenamefont {Scholtz}(2024)}]{Nandi:2023tfq}%
  \BibitemOpen
  \bibfield  {author} {\bibinfo {author} {\bibfnamefont {P.}~\bibnamefont {Nandi}}\ and\ \bibinfo {author} {\bibfnamefont {F.~G.}\ \bibnamefont {Scholtz}},\ }\href@noop {} {\bibfield  {journal} {\bibinfo  {journal} {Annals of Physics}\ }\textbf {\bibinfo {volume} {464}},\ \bibinfo {pages} {169643} (\bibinfo {year} {2024})}\BibitemShut {NoStop}%
\bibitem [{\citenamefont {Arzano}\ and\ \citenamefont {Kuipers}(2025)}]{Arzano:2025}%
  \BibitemOpen
  \bibfield  {author} {\bibinfo {author} {\bibfnamefont {M.}~\bibnamefont {Arzano}}\ and\ \bibinfo {author} {\bibfnamefont {F.}~\bibnamefont {Kuipers}},\ }\href@noop {} {\bibfield  {journal} {\bibinfo  {journal} {Physical Review D}\ }\textbf {\bibinfo {volume} {111}},\ \bibinfo {pages} {025010} (\bibinfo {year} {2025})}\BibitemShut {NoStop}%
\bibitem [{\citenamefont {Nandi}\ \emph {et~al.}(2025)\citenamefont {Nandi}, \citenamefont {Bhattacharyya}, \citenamefont {Majumdar}, \citenamefont {Pleasance},\ and\ \citenamefont {Petruccione}}]{Nandi:2025qyj}%
  \BibitemOpen
  \bibfield  {author} {\bibinfo {author} {\bibfnamefont {P.}~\bibnamefont {Nandi}}, \bibinfo {author} {\bibfnamefont {T.}~\bibnamefont {Bhattacharyya}}, \bibinfo {author} {\bibfnamefont {A.~S.}\ \bibnamefont {Majumdar}}, \bibinfo {author} {\bibfnamefont {G.}~\bibnamefont {Pleasance}},\ and\ \bibinfo {author} {\bibfnamefont {F.}~\bibnamefont {Petruccione}},\ }\href@noop {} {\bibfield  {journal} {\bibinfo  {journal} {arXiv preprint}\ } (\bibinfo {year} {2025})},\ \Eprint {https://arxiv.org/abs/2503.13061} {arXiv:2503.13061 [hep-th]} \BibitemShut {NoStop}%
\bibitem [{\citenamefont {Ashtekar}(1986)}]{ashtekar1986}%
  \BibitemOpen
  \bibfield  {author} {\bibinfo {author} {\bibfnamefont {A.}~\bibnamefont {Ashtekar}},\ }\href@noop {} {\bibfield  {journal} {\bibinfo  {journal} {Physical Review Letters}\ }\textbf {\bibinfo {volume} {57}},\ \bibinfo {pages} {2244} (\bibinfo {year} {1986})}\BibitemShut {NoStop}%
\bibitem [{\citenamefont {Breuer}\ and\ \citenamefont {Petruccione}(1992)}]{Breuer:1991ve}%
  \BibitemOpen
  \bibfield  {author} {\bibinfo {author} {\bibfnamefont {H.-P.}\ \bibnamefont {Breuer}}\ and\ \bibinfo {author} {\bibfnamefont {F.}~\bibnamefont {Petruccione}},\ }\href@noop {} {\bibfield  {journal} {\bibinfo  {journal} {Journal of Physics A}\ }\textbf {\bibinfo {volume} {25}},\ \bibinfo {pages} {L661} (\bibinfo {year} {1992})}\BibitemShut {NoStop}%
\bibitem [{\citenamefont {Breuer}\ \emph {et~al.}(1997)\citenamefont {Breuer}, \citenamefont {Kappler},\ and\ \citenamefont {Petruccione}}]{PhysRevA.56.2334}%
  \BibitemOpen
  \bibfield  {author} {\bibinfo {author} {\bibfnamefont {H.-P.}\ \bibnamefont {Breuer}}, \bibinfo {author} {\bibfnamefont {B.}~\bibnamefont {Kappler}},\ and\ \bibinfo {author} {\bibfnamefont {F.}~\bibnamefont {Petruccione}},\ }\href@noop {} {\bibfield  {journal} {\bibinfo  {journal} {Physical Review A}\ }\textbf {\bibinfo {volume} {56}},\ \bibinfo {pages} {2334} (\bibinfo {year} {1997})}\BibitemShut {NoStop}%
\bibitem [{\citenamefont {Breuer}\ and\ \citenamefont {Petruccione}(2007)}]{Breuer:2007juk}%
  \BibitemOpen
  \bibfield  {author} {\bibinfo {author} {\bibfnamefont {H.-P.}\ \bibnamefont {Breuer}}\ and\ \bibinfo {author} {\bibfnamefont {F.}~\bibnamefont {Petruccione}},\ }\href {https://doi.org/10.1093/acprof:oso/9780199213900.001.0001} {\emph {\bibinfo {title} {The Theory of Open Quantum Systems}}}\ (\bibinfo  {publisher} {Oxford University Press},\ \bibinfo {year} {2007})\BibitemShut {NoStop}%
\bibitem [{\citenamefont {Gaveau}\ \emph {et~al.}(1984)\citenamefont {Gaveau}, \citenamefont {Jacobson}, \citenamefont {Kac},\ and\ \citenamefont {Schulman}}]{gaveau1984}%
  \BibitemOpen
  \bibfield  {author} {\bibinfo {author} {\bibfnamefont {B.}~\bibnamefont {Gaveau}}, \bibinfo {author} {\bibfnamefont {T.}~\bibnamefont {Jacobson}}, \bibinfo {author} {\bibfnamefont {M.}~\bibnamefont {Kac}},\ and\ \bibinfo {author} {\bibfnamefont {L.~S.}\ \bibnamefont {Schulman}},\ }\href@noop {} {\bibfield  {journal} {\bibinfo  {journal} {Physical Review Letters}\ }\textbf {\bibinfo {volume} {53}},\ \bibinfo {pages} {419} (\bibinfo {year} {1984})}\BibitemShut {NoStop}%
\bibitem [{\citenamefont {Yordanov}(2024)}]{Yordanov:2024}%
  \BibitemOpen
  \bibfield  {author} {\bibinfo {author} {\bibfnamefont {V.}~\bibnamefont {Yordanov}},\ }\href@noop {} {\bibfield  {journal} {\bibinfo  {journal} {Scientific Reports}\ }\textbf {\bibinfo {volume} {14}},\ \bibinfo {pages} {6507} (\bibinfo {year} {2024})}\BibitemShut {NoStop}%
\bibitem [{\citenamefont {Rovelli}(2004)}]{rovelli2004}%
  \BibitemOpen
  \bibfield  {author} {\bibinfo {author} {\bibfnamefont {C.}~\bibnamefont {Rovelli}},\ }\href@noop {} {\emph {\bibinfo {title} {Quantum Gravity}}}\ (\bibinfo  {publisher} {Cambridge University Press},\ \bibinfo {year} {2004})\BibitemShut {NoStop}%
\bibitem [{\citenamefont {Alvarez-Gaume}\ and\ \citenamefont {Witten}(1984)}]{Alvarez-Gaume:1983ihn}%
  \BibitemOpen
  \bibfield  {author} {\bibinfo {author} {\bibfnamefont {L.}~\bibnamefont {Alvarez-Gaume}}\ and\ \bibinfo {author} {\bibfnamefont {E.}~\bibnamefont {Witten}},\ }\href {https://doi.org/10.1016/0550-3213(84)90066-X} {\bibfield  {journal} {\bibinfo  {journal} {Nucl. Phys. B}\ }\textbf {\bibinfo {volume} {234}},\ \bibinfo {pages} {269} (\bibinfo {year} {1984})}\BibitemShut {NoStop}%
\bibitem [{\citenamefont {Arnowitt}\ \emph {et~al.}(1962{\natexlab{b}})\citenamefont {Arnowitt}, \citenamefont {Deser},\ and\ \citenamefont {Misner}}]{ADM1962}%
  \BibitemOpen
  \bibfield  {author} {\bibinfo {author} {\bibfnamefont {R.}~\bibnamefont {Arnowitt}}, \bibinfo {author} {\bibfnamefont {S.}~\bibnamefont {Deser}},\ and\ \bibinfo {author} {\bibfnamefont {C.~W.}\ \bibnamefont {Misner}},\ }in\ \href@noop {} {\emph {\bibinfo {booktitle} {Gravitation: An Introduction to Current Research}}},\ \bibinfo {editor} {edited by\ \bibinfo {editor} {\bibfnamefont {L.}~\bibnamefont {Witten}}}\ (\bibinfo  {publisher} {Wiley},\ \bibinfo {address} {New York},\ \bibinfo {year} {1962})\ pp.\ \bibinfo {pages} {227--265},\ \bibinfo {note} {reprinted as arXiv:gr-qc/0405109}\BibitemShut {NoStop}%
\bibitem [{\citenamefont {Kuchař}(1992)}]{Kuchar1992}%
  \BibitemOpen
  \bibfield  {author} {\bibinfo {author} {\bibfnamefont {K.~V.}\ \bibnamefont {Kuchař}},\ }in\ \href@noop {} {\emph {\bibinfo {booktitle} {Proceedings of the 4th Canadian Conference on General Relativity and Relativistic Astrophysics}}},\ \bibinfo {editor} {edited by\ \bibinfo {editor} {\bibfnamefont {G.}~\bibnamefont {Kunstatter}}, \bibinfo {editor} {\bibfnamefont {D.}~\bibnamefont {Vincent}},\ and\ \bibinfo {editor} {\bibfnamefont {J.}~\bibnamefont {Williams}}}\ (\bibinfo  {publisher} {World Scientific},\ \bibinfo {address} {Singapore},\ \bibinfo {year} {1992})\ pp.\ \bibinfo {pages} {211--314}\BibitemShut {NoStop}%
\bibitem [{\citenamefont {Isham}(1992)}]{Isham1992}%
  \BibitemOpen
  \bibfield  {author} {\bibinfo {author} {\bibfnamefont {C.~J.}\ \bibnamefont {Isham}},\ }\href@noop {} {\bibinfo {title} {Canonical quantum gravity and the problem of time}},\ \bibinfo {howpublished} {Lecture at NATO Advanced Study Institute, Salamanca, 1992} (\bibinfo {year} {1992}),\ \Eprint {https://arxiv.org/abs/gr-qc/9210011} {gr-qc/9210011} \BibitemShut {NoStop}%
\bibitem [{\citenamefont {Page}\ and\ \citenamefont {Wootters}(1983)}]{Page:1983uc}%
  \BibitemOpen
  \bibfield  {author} {\bibinfo {author} {\bibfnamefont {D.~N.}\ \bibnamefont {Page}}\ and\ \bibinfo {author} {\bibfnamefont {W.~K.}\ \bibnamefont {Wootters}},\ }\href@noop {} {\bibfield  {journal} {\bibinfo  {journal} {Physical Review D}\ }\textbf {\bibinfo {volume} {27}},\ \bibinfo {pages} {2885} (\bibinfo {year} {1983})}\BibitemShut {NoStop}%
\end{thebibliography}%


\begin{thebibliography}{99}

\bibitem{Lindgren:2019tdd}
J.~Lindgren and J.~Liukkonen,
"Quantum Mechanics can be understood through stochastic optimization on spacetimes",
\textit{Scientific Reports} \textbf{9}, 19984 (2019).

\bibitem{fenyes1966}
I.~Fényes,
"Eine stochasticische Formulierung der quantenmechanischen Bewegungsgleichungen",
\textit{Zeitschrift für Physik} \textbf{132}, 81 (1966).

\bibitem{kershaw1966}
F.~Kuipers,
\textit{Stochastic Mechanics: The Unification of Quantum Mechanics with Brownian Motion},
arXiv:2301.05467 [quant-ph] (2023),
doi:10.1007/978-3-031-31448-3.



\bibitem{nelson1966}
E.~Nelson,
"Derivation of the Schrödinger equation from Newtonian mechanics",
\textit{Physical Review} \textbf{150}, 1079 (1966).

\bibitem{ambjorn2009}
J.~Ambjørn, R.~Loll, W.~Westra, and S.~Zohren,
"The emergence of (3+1)D spacetime in causal dynamical triangulations",
\textit{Physics Letters B} \textbf{678}, 1 (2009).

\bibitem{Unruh:1989db}
W.~G.~Unruh and R.~M.~Wald,
"Time and the interpretation of canonical quantum gravity",
\textit{Physical Review D} \textbf{40}, 2598--2614 (1989).

\bibitem{gaveau1984}
B.~Gaveau, T.~Jacobson, M.~Kac, and L.~S.~Schulman,
"Relativistic Extension of the Analogy Between Quantum Mechanics and Brownian Motion",
\textit{Physical Review Letters} \textbf{53}, 419--422 (1984).

\bibitem{ashtekar1986}
A.~Ashtekar,
"New variables for classical and quantum gravity",
\textit{Physical Review Letters} \textbf{57}, 2244 (1986).

\bibitem{arnowitt1962}
R.~Arnowitt, S.~Deser, and C.~W.~Misner,
"The Dynamics of General Relativity",
in \textit{Gravitation: An Introduction to Current Research}, edited by L.~Witten
(Wiley, New York, 1962) pp. 227--265.

\bibitem{Nandi:2025qrk}
P.~Nandi and P.~Ghose, "Stochastic Quantization of Electrodynamics and Linearized Gravity"
,
arXiv:2508.10190 [gr-qc] (2025).

\bibitem{Nandi:2023tfq}
P.~Nandi and F.~G.~Scholtz,
"The hidden Lorentz covariance of quantum mechanics",
\textit{Annals of Physics} \textbf{464}, 169643 (2024).

\bibitem{Arzano:2025}
M.~Arzano and F.~Kuipers,
"Stochastic origin of spacetime noncommutativity",
\textit{Physical Review D} \textbf{111}, 025010 (2025).

\bibitem{Nandi:2025qyj}
P.~Nandi, T.~Bhattacharyya, A.~S.~Majumdar, G.~Pleasance, and F.~Petruccione,
"Neutrino Decoherence in kappa-Minkowski Quantum Spacetime: An Open Quantum Systems Paradigm",
arXiv:2503.13061 [hep-th] (2025).

\bibitem{Breuer:1991ve}
H.-P.~Breuer and F.~Petruccione,
"Burgers' model of turbulence as a stochastic process",
\textit{Journal of Physics A} \textbf{25}, L661--L668 (1992).

\bibitem{PhysRevA.56.2334}
H.-P.~Breuer, B.~Kappler, and F.~Petruccione,
"Stochastic wave function approach to the calculation of multitime correlation functions of open quantum systems",
\textit{Physical Review A} \textbf{56}, 2334--2351 (1997).

\bibitem{Breuer:2007juk}
H.-P.~Breuer and F.~Petruccione,
\textit{The Theory of Open Quantum Systems} (Oxford University Press, 2007).

\bibitem{ashtekar1987}
A.~Ashtekar,
"New Hamiltonian formulation of general relativity",
\textit{Physical Review D} \textbf{36}, 1587 (1987).

\bibitem{Yordanov:2024}
V.~Yordanov,
"Derivation of Dirac equation from the stochastic optimal control principles of quantum mechanics",
\textit{Scientific Reports} \textbf{14}, 6507 (2024).

\bibitem{rovelli2004}
C.~Rovelli,
\textit{Quantum Gravity} (Cambridge University Press, 2004).

\bibitem{thiemann2007}
T.~Thiemann,
\textit{Modern Canonical Quantum General Relativity} (Cambridge University Press, 2007).

\bibitem{wheeler1968}
J.~A.~Wheeler,
"Superspace and the nature of quantum geometrodynamics",
in \textit{Battelle Rencontres: 1967 Lectures in Mathematics and Physics}, edited by C.~DeWitt and J.~A.~Wheeler (Benjamin, New York, 1968).

\bibitem{dewitt1967}
B.~S.~DeWitt,
"Quantum Theory of Gravity. I. The Canonical Theory",
\textit{Physical Review} \textbf{160}, 1113 (1967).

\bibitem{rovelli1995}
C.~Rovelli and L.~Smolin,
"Spin networks and quantum gravity",
\textit{Physical Review D} \textbf{52}, 5743 (1995).

\bibitem{perez2013}
A.~Perez,
"The Spin-Foam Approach to Quantum Gravity",
\textit{Living Reviews in Relativity} \textbf{16}, 3 (2013).

\bibitem{bojowald2005}
M.~Bojowald,
"Loop quantum cosmology",
\textit{Living Reviews in Relativity} \textbf{8}, 11 (2005).

\bibitem{gambini2005}
R.~Gambini and J.~Pullin,
\textit{A First Course in Loop Quantum Gravity} (Oxford University Press, 2011).

\bibitem{ADM1962}
R.~Arnowitt, S.~Deser, and C.~W.~Misner,
"The Dynamics of General Relativity",
in \textit{Gravitation: An Introduction to Current Research}, edited by L.~Witten (Wiley, New York, 1962), reprinted as arXiv:gr-qc/0405109.

\bibitem{Alvarez-Gaume:1983ihn}
L.~Alvarez-Gaumé and E.~Witten,
"Gravitational Anomalies",
\textit{Nucl. Phys. B} \textbf{234}, 269--330 (1984).


\bibitem{Kuchar1992}
K.~V.~Kuchař,
"Time and interpretations of quantum gravity",
in \textit{Proceedings of the 4th Canadian Conference on General Relativity and Relativistic Astrophysics}, edited by G.~Kunstatter, D.~Vincent, and J.~Williams (World Scientific, Singapore, 1992) pp. 211--314.

\bibitem{Isham1992}
C.~J.~Isham,
"Canonical quantum gravity and the problem of time",
Lecture at NATO Advanced Study Institute, Salamanca, 1992, arXiv:gr-qc/9210011 (1992).

\bibitem{Page:1983uc}
D.~N.~Page and W.~K.~Wootters,
"Evolution without evolution: Dynamics described by stationary observables",
\textit{Physical Review D} \textbf{27}, 2885--2892 (1983).


\end{thebibliography}

\end{document}